\documentclass[10pt]{iopart}

\usepackage{iopams}  
\usepackage{graphicx}

\begin{document}

\title{Comments on ``Ohm's Law Survives to the Atomic Scale'' by Weber et al.
}

\author{Mukunda P. Das$^{1*}$ and Frederick Green$^2$}
\address{$^1$ Department of Theoretical Physics,
Research School of Physics and Engineering,
The Australian National University, Canberra, ACT 0200, Australia.}
\address{$^2$ School of Physics, The University of New South Wales,
Sydney, NSW 2052, Australia.}
\address{$^*$ Corresponding author: {\em mukunda.das@anu.du.au}}

\begin{abstract}
The recent article "Ohm's Law Survives to the Atomic Scale" by Weber et al.
\cite{1}
reveals ohmic transport in quantized P-in-Si wires. We argue that their
results have two main deficiencies: (a) the interpretation of conductance
data is inadequate for serious systematics; (b) metallic-like structures hold
few implications for quantum computing
\cite{0}.
\end{abstract}

In their recent paper Weber et al. \cite{1} claim novel observations on the
persistence of ohmic behavior in embedded P-in-Si wires, effectively
fabricated at atomic scales. To further promote the sophisticated and
quite delicate materials processing achieved in this work, it would be
important to retain certain theoretical
issues -- issues of principle -- clearly to the fore. To that end we
offer below several comments on the authors' theoretical interpretation
of their observed results. 

\smallskip
{\bf 1}. The statement that
``The unambiguous demonstration of Ohmic scaling is a
constant resistivity $\rho_{\rm W}= R_{\rm W}(A_{\rm el}/L)$ independent of geometric variables,
such as wire length or width'' does not appear thoroughly conclusive from the
published data. Table 1 and Fig 1E contain the main results of this work.  In
Table 1, $\rho_{\rm W}$, which by convention defines an {\em intrinsic}
material constant,
varies between 0.10 and 0.43 m$\Omega$cm. Fig. 1E presents quite sparse and
scattered data for the product $\rho_{\rm W} w$, with $w$ the nominal wire width,
in which the linear interpolation fails to cut the stated error bars.
The plot is log-log; once presented on the linear scale, the real
magnitudes of these discrepancies are much more evident. It is hard to
reasonably describe the claimed ohmic relationship as ``unambiguous''.

\smallskip
{\bf 2}. The STM-patterned P structure in a Si matrix constitutes a
quasi-one-dimensional (1D) wire of high (metallic) carrier density. In
Fig. 2C, the presented data for $\rho_{\rm W}$ are said to correspond to bulk
electron densities of $10^{21}{\rm cm}^{-3}$.
The physics in actual question here concerns {\em one-dimensional} transport.
Thus, what is the physical relevance of a bulk electronic density
in the 1D context? The current-carrying capacity of P-in-Si nanowires is
further compared with a nominal Cu equivalent,
scaled to comparable dimensions.
For Cu at $T=4.2$K, we have $\rho_{\rm W} \sim 1 \mu\Omega{\rm cm}$ while,
for the P-in-Si wires, $\rho_{\rm W}$ comes out $\sim 0.1 {\rm m}\Omega{\rm cm}$
This is a difference of two orders of magnitude.    

\smallskip
{\bf 3}. The (finite) resistance of these P-in-Si wires is in the
highly diffusive regime and, hence, far from ballistic ideality.
The mean free path is $l \sim 8±1$nm. According to the Landauer
formula in its diffusive elastic regime
\cite{2},
$R_{\rm calc} = (h/2e2N) (1+L/l)$. This is a two-terminal formula,
which consequently includes both the true wire {\em and}
its contact resistances
\cite{2}. To cite just one example: wire W1 has $L=312$nm,
$w=11.0$nm and $N \approx 20$, yielding $R_{\rm calc}=25.8{\rm k}\Omega$.
This presents a large resistance
leading to correspondingly large values of local Joule heating, at
least in the voltage ranges adopted for industrial-standard Si
microelectronics.     

\smallskip
{\bf 4}. The measured resistance, as stated in the caption of Fig.1E, is a
four-terminal result; that is, it should be the {\em intrinsic}
wire resistance, sans any contact resistance at all. On the other hand
the formula as applied for calculated resistance, quoted from Weber et al.
in our preceding Point 3, is a two-terminal result that
{\em includes} the contact resistance
\cite{2}.

The difference between these two distinct resistances in the ideal metallic
limit can be drastic, as attested by the definitive 1D wire
 measurements of dePicciotto et al.
\cite{3}.
In the work of Weber et al. the contact resistance works out at
$12.9/N {\rm k}\Omega$. This is much smaller than for the wire
itself, which is highly resistive. One can safely conclude that
the P-in-Si wire dominates the overall resistive behaviour.
In other words, it behaves ohmically because it is an unexceptionally
ohmic structure from the start; from the operational
standpoint, its ``atomic'' nature is irrelevant
(all metallic resistance is the direct outcome of strong processes
at the quantum level).

Aside from the above, there is a technical confusion.
The experimental $R_{\rm W}$ is a four-terminal result;
but $R_{\rm calc}$ follows the two-terminal formula (see their Table 1).
The authors claim good agreement (including wires 4 and 5)
between these two structurally very different theoretical quantities.
This means that the influence of the contacts (within the Landauer formula)
ought to be negligible. But for samples 4 and 5,
there is a large mismatch; thus the effect of the contact resistances becomes
hard to quantify.

Another feature to note, and hopefully understand, is that
the measured wire resistances are {\em higher}
than the calculated resistances for wire samples 1, 4 and 5
but {\em lower} than those for wires 2 and 3. In our view
the models used for analysis of the observations are inconsistent and
inadequate.

\smallskip
{\bf 5}. Finite resistance necessarily entails power dissipation.
Here, the heat
production would be $\sim\!\!100$ times that of a Cu channel of equivalent
geometry. The paper offers no analysis of the practical implications for
quantum-computer architectures, where the extreme feature densities of such
P-in-Si metallization would exacerbate (as a notorious side-effect of Moore's
Law) the problem of highly localized heat production and its removal. An
ohmic wire -- however made -- remains highly dissipative and thus a problematic
destroyer of quantum entanglement were it to be seriously considered,
as envisaged, for interconnects in quantum computing.


\section*{References}
\begin{thebibliography}{99}

\bibitem{1}
B. Weber et al., Science {\bf 335}, 64 (2012).

\bibitem{0}
This Comment has been released on the Science web site at\\
{\em http://comments.sciencemag.org/content/10.1126/science.1214319\#comments}

\bibitem{2}
M. J. M. de Jong and C. W. J. Beenakker, Phys. Rev. B {\bf 51}, 16867 (1995).

\bibitem{3}
R. dePicciotto, H. L. Stormer, L. N. Pfeiffer, K. W. Baldwin and
K. W. West, Nature {\bf 411}, 51 (2001).

\end{thebibliography}
\end{document}